\documentclass[review]{elsarticle}
\graphicspath{ {./figures/} }
\usepackage{hyperref}
\usepackage{float}
\usepackage{verbatim} 
\usepackage{amsmath, amssymb}
\usepackage{bm}
\usepackage{booktabs}
\usepackage{graphicx}
\restylefloat{figure}
\floatstyle{plaintop} 
\restylefloat{table}

\begin{document}

\begin{frontmatter}

\title{Smart Predict--then--Optimize Paradigm for Portfolio Optimization in Real Markets}

\author[inst1]{Yi Wang}
\ead{apple9238@fuji.waseda.jp}

\author[inst1]{Takashi Hasuike}
\ead{thasuike@waseda.jp}

\address[inst1]{Waseda University, Graduate School of Creative Science and Engineering, Tokyo, Japan}

\begin{abstract}
Improvements in return forecast accuracy do not always lead to proportional improvements in portfolio decision quality, especially under realistic trading frictions and constraints. This paper adopts
the Smart Predict--then--Optimize (SPO) paradigm for portfolio optimization in
real markets, which explicitly aligns the learning objective with downstream
portfolio decision quality rather than pointwise prediction accuracy. Within
this paradigm, predictive models are trained using an SPO-based surrogate loss
that directly reflects the performance of the resulting investment decisions. To preserve interpretability and robustness, we employ linear predictors built
on return-based and technical-indicator features and integrate them with
portfolio optimization models that incorporate transaction fee, turnover
control, and regularization. We evaluate the proposed approach on U.S. ETF data
(2015--2025) using a rolling-window backtest with monthly rebalancing. Empirical
results show that decision-focused training consistently improves risk-adjusted
performance over predict--then--optimize baselines and classical optimization
benchmarks, and yields strong robustness during adverse market regimes (e.g.,
the 2020 COVID-19). These findings highlight the practical value of the Smart
Predict--then--Optimize paradigm for portfolio optimization in realistic and
non-stationary financial environments.
\end{abstract}


\end{frontmatter}

\section{Introduction}

Portfolio optimization is a central problem in financial decision-making, where
investors seek to allocate capital across multiple assets to balance return and
risk. With the increasing availability of financial data, machine learning
techniques have been widely adopted to support data-driven portfolio
construction and asset allocation. Despite these advances, financial markets remain highly complex systems
characterized by nonlinearity, noise, and non-stationarity, making robust and
reliable investment decision-making challenging in practice.
These characteristics expose a fundamental tension between improving
predictive accuracy and achieving high-quality portfolio decisions under realistic market frictions.

A large body of prior work focuses on improving return predictability using
advanced time-series models, including recurrent neural networks and
Transformer-based architectures. While these models have demonstrated strong
performance in various forecasting tasks, their practical benefits in financial
applications are often limited by weak periodicity, low signal-to-noise ratios,
and frequent regime shifts. Moreover, recent evidence suggests that simple and
transparent models can achieve competitive predictive performance, calling into
question the necessity of highly complex architectures in noisy financial
environments. More importantly, improved forecasting accuracy alone does not
guarantee superior portfolio performance.

Most learning-based portfolio methods adopt a predict--then--optimize (PtO)
paradigm, in which asset returns or related parameters are first estimated and
then treated as fixed inputs to a downstream optimization problem. Although
intuitive, this decoupled approach suffers from a fundamental mismatch between
the training objective and the ultimate decision goal. In particular, small
prediction errors can induce large changes in optimal portfolio weights,
especially under realistic constraints and trading frictions, leading to
suboptimal investment decisions even when predictive accuracy is high.

To address this limitation, the Smart Predict--then--Optimize (SPO) framework
and, more broadly, decision-focused learning (DFL) paradigms have been proposed
to explicitly realign the learning objective with downstream decision quality.
From a paradigm perspective, SPO departs from the conventional
predict--then--optimize workflow by embedding the optimization problem into the
training loop, thereby training predictive models based on the quality of the
resulting decisions rather than prediction error alone.

In this study, we investigate decision-focused learning for portfolio
optimization under a realistic rolling-window backtesting framework. We adopt
linear predictors for interpretability and robustness, and integrate them with
multiple portfolio optimization formulations within a unified SPO-based
architecture. Our contributions can be summarized as follows:
\begin{itemize}
    \item We systematically evaluate the Smart Predict--then--Optimize paradigm for portfolio optimization under realistic trading constraints and transaction fee, and contrast it with conventional predict--then--optimize approaches.
    \item We propose robust extensions of the SPO framework that explicitly
    account for prediction uncertainty and demonstrate their effectiveness
    during adverse market regimes.
    \item Through extensive empirical studies on U.S. ETF data across multiple
    market conditions, including crisis and bull-market periods, we show that
    decision-focused learning consistently improves portfolio decision quality
    compared to conventional predict--then--optimize baselines.
\end{itemize}

\section{Literature Review}

\subsection{Portfolio Optimization Modeling}

Modern portfolio optimization originates from the mean--variance framework
proposed by Markowitz~\cite{markowitz1952portfolio}, which formulates asset
allocation as a trade-off between expected return and risk.
Subsequent studies have extended this framework by introducing alternative
risk measures, additional constraints, and techniques to address estimation
uncertainty, including the Sharpe ratio~\cite{sharpe1998sharpe} and
coherent risk measures such as Conditional Value-at-Risk (CVaR)~\cite{Rockafellar2000Cvar}.
Robust portfolio optimization further accounts for parameter uncertainty by
optimizing portfolio decisions against worst-case scenarios within prescribed
uncertainty sets~\cite{goldfarb2003robust}.

From both theoretical and empirical perspectives, Jagannathan and Ma~\cite{Jagannathan2003constraints}
demonstrate that portfolio weight constraints can be interpreted as a form of
statistical regularization.
Such constraints mitigate estimation error and improve out-of-sample risk
performance, thereby providing a methodological foundation for incorporating
additional constraints and penalty terms in portfolio optimization models.

In particular, transaction fee and portfolio turnover are commonly modeled
through penalty terms in the optimization objective to account for realistic
trading frictions~\cite{Boyd2017MultiPeriodTV}.

\subsection{Learning-based Approaches}

Learning-based approaches to portfolio optimization can be broadly divided into two paradigms according to how learning and decision-making are integrated: reinforcement learning (RL) methods and predict--then--optimize (PtO) methods. 

Reinforcement learning approaches formulate portfolio management as a sequential decision-making problem under a Markov decision process, where an agent learns a policy to maximize cumulative rewards through repeated interaction with a financial environment. A variety of RL~\cite{J.Moody2001ReccurentRL} and deep reinforcement learning (DRL)~\cite{Jiang2017DRL} algorithms have been applied in this context, including DQN~\cite{Pigorsch2022DQL}, PPO~\cite{Liu2024PPO}, and DDPG~\cite{Deng2017RepresentDRL}. While these approaches offer high modeling flexibility, they typically rely on carefully designed reward functions and environment simulators, which may lead to unstable training dynamics and limited interpretability in noisy and non-stationary financial markets~\cite{Bai2025RLReview}.

In contrast, predict--then--optimize (PtO) methods decouple the prediction and decision stages by first learning problem parameters, such as asset returns or costs, and then solving a downstream optimization problem to obtain portfolio decisions. This paradigm preserves the structure of classical portfolio optimization models and enables the explicit incorporation of constraints and domain knowledge.

Within the PtO framework, a wide variety of machine learning and neural network models have been extensively investigated for portfolio optimization. In these approaches, learning models are used to estimate key problem parameters, including expected asset returns, risk measures, or other predictive signals, which are subsequently incorporated into classical portfolio optimization formulations. Representative predictive models include linear regression~\cite{DengMin2013LinearforPtO}, support vector machine(SVM)~\cite{PAIVA2019SVMforPtO}, tree-based method, and deep neural network (DNN)~\cite{Freitas2009PredictionBasedPtO}, such as recurrent neural network (RNN) and long short-term memory (LSTM)~\cite{WANG202011LSTMforPtO} network. Comprehensive reviews and empirical studies ~\cite{KRAUSS2017Review1, MA2021Review2}have examined the effectiveness of these models in portfolio-related applications. By integrating data-driven predictions with optimization-based portfolio construction, PtO methods aim to improve investment performance while maintaining the interpretability and structural advantages of traditional portfolio optimization frameworks.

However, stock market prediction remains a challenging time-series forecasting problem, as financial markets are inherently nonlinear, dynamic, noisy, and chaotic, making direct and accurate prediction difficult~\cite{Deboeck1994difficulty}.

Recent advances in recurrent neural networks such as LSTM~\cite{Lindemann2021LSTMsurvey} and Transformer-based models~\cite{Wang2024TimeXer, Zhang2023TimeMixer, Nie2023PatchTST, Liu2023iTransformer} have demonstrated strong effectiveness across a wide range of time-series forecasting tasks. Nevertheless, these architectures often rely on inductive biases such as long-range dependency modeling or periodic pattern extraction, which are particularly suitable for structured or seasonal time series. In contrast, financial time series are characterized by weak periodicity, low signal-to-noise ratios, and frequent regime shifts, limiting the practical gains of highly complex architectures in real-world investment settings.

Moreover, the increased architectural complexity of deep recurrent and Transformer-based models raises concerns regarding training stability and interpretability in high-stakes financial decision-making. For example, recent studies such as DLinear~\cite{Zeng2021DLinear} question the necessity of complex architectures for time-series forecasting, showing that simple linear models can achieve competitive performance.

Nevertheless, even with simplified and more transparent predictive models, the relationship between prediction accuracy and downstream portfolio decision quality remains indirect. Under these circumstances, conventional predict--then--optimize (PtO) approaches that primarily focus on predictive accuracy may fail to yield optimal investment decisions. This limitation motivates the adoption of Smart Predict--then--Optimize (SPO) and other decision-focused learning frameworks, which explicitly align the learning objective with downstream portfolio optimization performance.

\subsection{Smart Predict-then-Optimize (SPO)}

The distinction between predict--then--optimize (PtO) and Smart Predict--then--Optimize (SPO) was formally introduced by Elmachtoub and Grigas~\cite{elmachtoub2022SPO}. While PtO methods focus on minimizing prediction error independently of the downstream optimization problem, SPO explicitly aligns the learning objective with decision quality. In related literature, this paradigm is also referred to as prediction-focused learning and decision-focused learning~\cite{mandi2024DFLReview}.

In a conventional Predict-then-Optimize (PtO) framework, the predictive model is trained by minimizing a surrogate loss function, such as the mean squared error between predicted and realized parameters. The resulting predictions are then treated as fixed inputs to a downstream optimization problem. While this approach is intuitive and computationally convenient, minimizing prediction error does not necessarily lead to high-quality decisions.

This misalignment arises because small prediction errors can result in large changes in the optimal solution of the optimization problem, especially in constrained or ill-conditioned settings. Consequently, improvements in predictive accuracy do not always translate into better downstream decision quality.

To address this issue, the Smart Predict-then-Optimize (SPO) framework proposes to directly incorporate the downstream optimization problem into the learning objective. Instead of optimizing prediction accuracy alone, SPO minimizes a decision-oriented loss, typically defined in terms of decision regret, thereby explicitly aligning the training objective with the quality of the resulting portfolio decisions.

This perspective is especially relevant in financial applications. As discussed earlier, stock return prediction is inherently difficult due to market noise, non-stationarity, and complex dynamics, making prediction errors inevitable. At the same time, financial decision-making places stringent requirements on robustness, stability, and risk control, where small prediction errors may lead to disproportionately large economic losses. Under such conditions, learning paradigms that explicitly optimize decision quality rather than pointwise prediction accuracy are particularly valuable. Recent work has further emphasized the importance of robustness in decision-focused learning by proposing regret-based loss functions that explicitly hedge against adverse decision outcomes~\cite{schutte2024robust}, motivating the adoption of robust SPO-based approaches for portfolio optimization.

Preliminary ideas related to this study were explored in an earlier conference version~\cite{Wang2025EndToEndSPO}. The present work substantially extends this line of research by providing a systematic evaluation and more comprehensive modeling. Building upon this initial investigation, the present study conducts a more systematic evaluation of SPO-based methods in portfolio optimization.

In addition, this study adopts PyEPO~\cite{tang2024pyepo} to implement the SPO loss. PyEPO offers a convenient and flexible framework for SPO modeling and has been increasingly used in recent working papers and preprints.

\section{Methodology}
\subsection{Smart Predict then Optimize (SPO) based model}

In this study, we adopt a Smart Predict--then--Optimize (SPO) framework for portfolio optimization, where prediction and decision-making are tightly coupled through an optimization layer.
Due to the computational and implementation complexity associated with backpropagation through optimization problems, we focus on tractable yet expressive portfolio formulations within a unified decision-focused learning architecture.

To preserve model interpretability and adhere to the principle of Occam’s razor, we consistently employ linear models as predictors across all experimental settings. Linear predictors provide a transparent mapping from input features to asset return estimates, which facilitates interpretability and contributes to improved robustness of the resulting portfolio decisions.

\begin{figure}[H] 
\centering 
\includegraphics[width=1.0\linewidth]{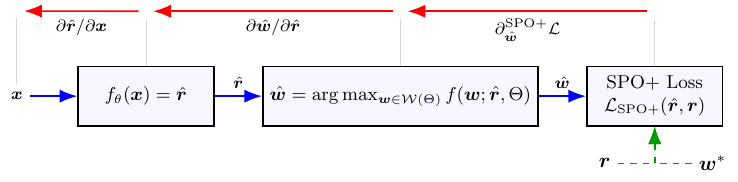} 
\caption{Overview of the Smart Predict--then--Optimize (SPO) training pipeline for portfolio optimization. } 
\label{fig:SPO+} 
\end{figure}

Figure~\ref{fig:SPO+} illustrates the overall predictive--optimization pipeline.
Given input features $\bm{x}$, a predictor produces estimated returns $\hat{\bm{r}}$, which are then passed to a parameterized portfolio optimization module to obtain the portfolio weights.
The resulting decision is evaluated using the SPO+ loss, enabling end-to-end training by propagating gradients through both the prediction model and the optimization layer.

Within this unified structure, different portfolio optimization models are instantiated by specifying different objective functions and optimization parameters, collectively denoted by $\Theta$.
These instantiations reflect different investment objectives and practical trading considerations.
In the following, we consider three representative formulations: \textbf{MaxReturn}, \textbf{MaxReturn with transaction fee}, and \textbf{MaxReturn with $\ell_2$ weight regularization}.

\subsubsection{MaxReturnSPO}

We first consider the basic MaxReturn formulation, which serves as the simplest instantiation of the SPO framework and forms the foundation for subsequent extensions.

Given an input feature vector $\bm{x} \in \mathbb{R}^d$, a trainable predictor $f_\theta(\cdot)$ produces an estimate of the asset return vector:
\begin{equation}
\hat{\bm{r}} = f_\theta(\bm{x}).
\end{equation}

Based on the predicted returns, the portfolio decision is obtained by solving a return-maximization problem:
\begin{equation}
\hat{\bm{w}} = \arg\max_{\bm{w} \in \mathcal{W}} \hat{\bm{r}}^\top \bm{w},
\end{equation}
where $\mathcal{W}$ denotes the feasible set of portfolio weights.

During training, the realized return vector $\bm{r}$ is assumed to be observable, and the corresponding oracle decision is defined as
\begin{equation}
\bm{w}^\star = \arg\max_{\bm{w} \in \mathcal{W}} \bm{r}^\top \bm{w}.
\end{equation}

To enable end-to-end learning, we adopt the SPO+ surrogate loss, which provides a convex upper bound on the true decision regret:
\begin{equation}
\mathcal{L}_{\mathrm{SPO+}}(\hat{\bm{r}}, \bm{r})
=
\max_{\bm{w} \in \mathcal{W}}
(2\hat{\bm{r}} - \bm{r})^\top \bm{w}
-
\bm{r}^\top \bm{w}^\star.
\end{equation}

\subsubsection{MaxReturn with transaction fee}
To account for practical trading frictions, we extend the MaxReturnSPO framework by explicitly incorporating transaction fee into the portfolio optimization objective. Let $\bm{w}_{t-1}$ denote the portfolio held in the previous period. The decision variable is obtained by solving
\begin{equation}
\hat{\bm{w}}
=
\arg\max_{\bm{w} \in \mathcal{W}}
\left(
\hat{\bm{r}}^\top \bm{w}
-
\gamma \lVert \bm{w} - \bm{w}_{t-1} \rVert_1
\right),
\end{equation}
where $\gamma > 0$ controls the transaction fee intensity.\\

\noindent\textbf{Notation.}
Let $\hat{\bm r}\in\mathbb{R}^n$ denote the predicted return vector,
$\bm r\in\mathbb{R}^n$ the realized return vector,
$\bm w\in\mathbb{R}^n$ the portfolio weight vector,
and $\mathcal W$ the feasible set of portfolios.
Let $\bm w_{t-1}$ denote the portfolio held in the previous period and
$\gamma>0$ the transaction-fee coefficient.
We define the value function
\[
f(\hat{\bm r}) = \max_{\bm w\in\mathcal W}
\left[(\hat{\bm r}-\bm r)^\top \bm w
- \gamma \|\bm w-\bm w_{t-1}\|_1 \right].
\]
By Danskin's theorem, if
$\tilde{\bm w}\in\arg\max_{\bm w\in\mathcal W}(\cdot)$,
then $\tilde{\bm w}\in\partial_{\hat{\bm r}} f(\hat{\bm r})$.

\subsubsection{MaxReturn with $\ell_2$ Weight Regularization}

We observe that the previous formulations may yield highly concentrated portfolios, occasionally allocating most of the capital to a single asset. Such extreme allocations are often undesirable in practice due to their instability and sensitivity to prediction errors. To alleviate this issue, we introduce an $\ell_2$ regularization term on portfolio weights, which penalizes large positions and encourages more balanced and stable allocations.

Specifically, the portfolio decision is obtained by solving
\begin{equation}
\label{eq:maxreturn_l2_reg}
\hat{\bm w}
=
\arg\max_{\bm w\in\mathcal W}
\left(
\hat{\bm r}^\top \bm w
-
\gamma \|\bm w-\bm w_{t-1}\|_1
-
\lambda \|\bm w\|_2^2
\right),
\end{equation}
where $\hat{\bm r}\in\mathbb{R}^n$ denotes the predicted return vector,
$\bm w\in\mathbb{R}^n$ the portfolio weight vector,
$\mathcal W$ the feasible set of portfolios,
$\bm w_{t-1}$ the portfolio held in the previous period,
$\gamma>0$ the transaction fee coefficient, and
$\lambda>0$ the regularization parameter controlling the strength of the $\ell_2$ penalty.\\
Since the $\ell_2$ regularization term does not depend on the predicted returns, the resulting optimization problem remains affine in $\hat{\bm r}$, and gradients can still be propagated through the optimization layer using the SPO+ surrogate.

\subsubsection{RobustSPO}

While the standard SPO framework optimizes portfolio decisions based on point
predictions of asset returns, such predictions are inevitably subject to
estimation errors.
To improve robustness against adverse prediction perturbations, we further
extend the MaxReturnSPO formulation by incorporating robustness into the
prediction--decision pipeline.

Specifically, we consider multiplicative perturbations applied to the predicted
return vector.
Let $\boldsymbol{\zeta}\in\mathbb{R}^n$ denote a perturbation vector and define
the perturbed prediction as
\begin{equation}
\tilde{\bm r}
=
\hat{\bm r} \circ (1+\boldsymbol{\zeta}),
\end{equation}
where $\circ$ denotes the element-wise product.
The perturbation is assumed to lie in the uncertainty set
\begin{equation}
\mathcal U
=
\left\{
\boldsymbol{\zeta} :
\|\boldsymbol{\zeta}\|_\infty \le \rho
\right\},
\end{equation}
where $\rho>0$ controls the maximum relative deviation of each asset return.

Given a perturbed prediction $\tilde{\bm r}$, the corresponding portfolio
decision is obtained by solving
\begin{equation}
\tilde{\bm w}
=
\arg\max_{\bm w\in\mathcal W}
\tilde{\bm r}^\top \bm w .
\end{equation}

The robust SPO objective is then formulated as a worst-case regret minimization
problem:
\begin{equation}
\min_{\theta}
\;
\max_{\boldsymbol{\zeta}\in\mathcal U}
\left(
\bm r^\top \bm w^\star
-
\bm r^\top \tilde{\bm w}
\right),
\end{equation}
where $\bm w^\star$ denotes the oracle portfolio under the realized return vector
$\bm r$.

In practice, the inner maximization over the uncertainty set is approximated via
Monte Carlo sampling.
For each mini-batch, multiple perturbed return predictions are generated, and
the maximum SPO+ loss among these scenarios is used for backpropagation.
This robust training procedure encourages the predictor to produce portfolio
decisions that are less sensitive to unfavorable prediction errors.

\subsection{SoftmaxDFL}
Motivated by prior work~\cite{Zhang2020DeepPortfolio} that proposes constructing
portfolio weights via a softmax layer and directly optimizing financial
performance metrics through a loss function, we also consider an alternative
learning-based portfolio allocation approach based on a fully differentiable
allocation structure.
As a differentiable baseline, we consider a neural allocator that directly maps
input features to portfolio weights via a softmax layer, denoted as
\textbf{SoftmaxDFL}.
Unlike SPO-based methods that predict returns and then solve an explicit
optimization problem, SoftmaxDFL learns the allocation policy end-to-end without
an explicit optimization layer.

\begin{figure}[H] 
\centering 
\includegraphics[width=1.0\linewidth]{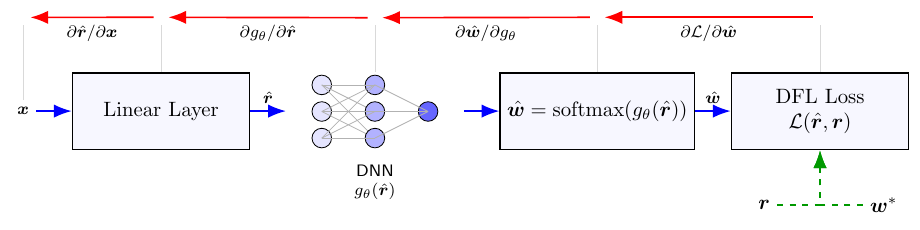} 
\caption{Architecture of the SoftmaxDFL. } 
\label{fig:SoftmaxDFL} 
\end{figure}

Given input features $\bm x$, a linear inferencer $f_\theta(\cdot)$ first produces
return estimates
\begin{equation}
\hat{\bm r} = f_\theta(\bm x).
\end{equation}
A neural allocator $g_\theta(\cdot)$ then maps $\hat{\bm r}$ directly to portfolio
weights via a softmax layer:
\begin{equation}
\hat{\bm w} = g_\theta(\hat{\bm r}),
\end{equation}
which ensures nonnegative weights that sum to one.
Unlike SPO-based methods, SoftmaxDFL does not solve an explicit optimization
problem in the forward pass.

The learned allocation is evaluated based on its realized portfolio performance.

\textbf{MaxReturn} optimizes portfolio performance by minimizing the negative
realized return:
\begin{equation}
\mathcal L_{\mathrm{MaxReturn}}
=
-\,\bm r^\top \hat{\bm w}.
\end{equation}

\textbf{MaxSharpe} considers risk-adjusted performance by minimizing a
differentiable surrogate of the Sharpe ratio:
\begin{equation}
\mathcal L_{\mathrm{MaxSharpe}}
=
-
\frac{\bm r^\top \hat{\bm w}}
{\sqrt{\hat{\bm w}^\top\Sigma \hat{\bm w}}},
\end{equation}
where $\Sigma$ denotes the return covariance matrix estimated from the training
window.

\subsection{Baselines}

We consider two classical baselines for portfolio optimization.

\subsubsection{Max Sharpe Ratio}
As a traditional optimization-based benchmark, we adopt the classical
mean--variance portfolio model.
Given the historical mean return vector $\bar{\bm r}$ and covariance matrix $\Sigma$,
the optimal portfolio is obtained by solving
\begin{equation}
\bm w^\star
=
\arg\max_{\bm w\in\mathcal W}
\frac{\bar{\bm r}^\top \bm w}
{\sqrt{\bm w^{\top}\Sigma \bm w}} .
\end{equation}

\subsubsection{Predict-then-Optimize (PtO)}

As a Predict-then-Optimize (PtO) baseline, we combines a linear return predictor with the classical Markowitz model.
Specifically, given input features $\bm x$, a linear model $f_\theta(\cdot)$ is trained
to predict asset returns by minimizing the mean squared error:
\begin{equation}
\hat{\bm r} = f_\theta(\bm x),
\qquad
\min_{\theta}\;
\|\hat{\bm r}-\bm r\|_2^2 .
\end{equation}

After training the predictor, the predicted returns are treated as fixed inputs
to the portfolio optimization stage.
The portfolio weights are then obtained by solving the Markowitz problem
\begin{equation}
\bm w^{\mathrm{PtO}}
=
\arg\max_{\bm w\in\mathcal W}
\hat{\bm r}^\top \bm w .
\end{equation}

This baseline explicitly decouples return prediction and portfolio optimization,
and serves as a standard comparison to decision-focused learning approaches.

\section{Experiment Setup}
\subsection{Data Description}
To evaluate the proposed SPO framework for portfolio optimization, we employ daily historical data of U.S. exchange-traded funds (ETFs) as the experimental dataset from 2015/01/01 to 2025/01/01. Each asset in the investment universe corresponds to a tradable ETF, and the dataset spans multiple consecutive years of trading days, enabling a realistic portfolio backtesting environment.

At each trading day, the model takes return-based and technical indicator features derived from historical prices as inputs to predict the next-period asset returns (or equivalently, a cost vector), which are then fed into a portfolio optimization model to generate portfolio allocation weights. Specifically, the input features include:\\
(1) \textbf{log returns}, capturing short-term price movements;\\
(2) \textbf{simple moving averages} (SMA) and \textbf{price bias}, reflecting trend-following and mean-reversion characteristics;\\
(3) \textbf{relative strength index} (RSI) and \textbf{MACD differences}, capturing momentum and trend dynamics;\\
(4) \textbf{Bollinger band width}, measuring market volatility; \\
(5) \textbf{volume-based indicators}, reflecting trading activity and market participation.

These features jointly describe the market state of each asset from the perspectives of returns, trends, momentum, volatility, and trading behavior, providing economically meaningful inputs for the prediction models. This data setting naturally aligns with the decision-focused learning paradigm, where prediction quality is evaluated through downstream portfolio decisions and backtesting performance rather than pointwise forecasting accuracy.

\subsection{Backtest Strategy}

We evaluate all models using a rolling-window backtesting framework with monthly rebalancing, which reflects a realistic investment process. At each rebalancing month $t$, the prediction model is retrained from scratch using only historical information available prior to $t$.

\begin{figure}[H] 
\centering 
\includegraphics[width=1.0\linewidth]{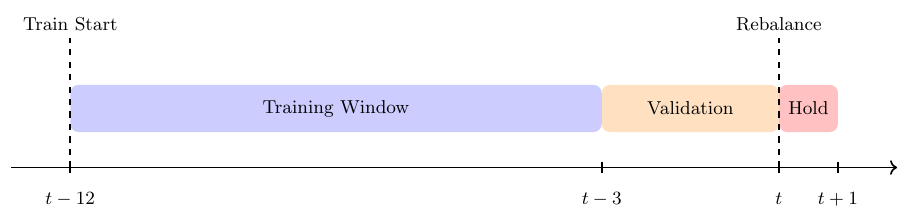} 
\caption{Rolling-window backtesting protocol with monthly rebalancing and time-series validation.} 
\label{fig:BacktestStrategy} 
\end{figure}

As illustrated in Figure ~\ref{fig:BacktestStrategy}, for each month $t$, we construct a fixed-length training window covering the past 12 months $[t-12,\,t-3]$, followed by a time-ordered validation period $[t-3,\,t]$. The window is rolled forward by one month at each rebalancing step.

Within each rolling window, hyperparameters are selected via \textbf{time-series validation} by Optuna, preserving the temporal order of the data. After hyperparameter selection, the model is trained on the corresponding window and used to generate predicted return vectors, which serve as inputs to the portfolio optimization problem. The resulting portfolio is held over the out-of-sample period $[t,\,t+1]$ until the next rebalancing date.

In all backtesting experiments, a proportional transaction fee of 0.005 is deducted
from portfolio returns based on portfolio turnover, measured by
$\|\bm w_t - \bm w_{t-1}\|_1$, to account for realistic trading frictions.
This transaction fee is applied uniformly to all compared strategies.
\subsection{Model Configuration}

\begin{table}[H]
\centering
\caption{Overview of compared strategies.}
\label{tab:strategy_mapping}
\begin{tabular}{ll}
\toprule
\textbf{Category} & \textbf{Strategies} \\
\midrule
SoftmaxDFL
& Softmax-MaxReturn, Softmax-MaxSharpe \\

Robust SPO
& RobustSPO ($\rho=0.01$), RobustSPO ($\rho=0.1$) \\

SPO
& SPO+, SPO+ with Fee, SPO+ with turnover penalty \\

PtO baseline
& PtO Markowitz \\

Classical baseline
& MaxSharpe \\
\bottomrule
\end{tabular}
\end{table}

\begin{table}[H]
\centering
\caption{Model Configuration of SPO models}
\label{tab:SPO_params}
\begin{tabular}{ll}
\hline
\textbf{Parameter} & \textbf{Values} \\
\hline
Proportional transaction fee $\gamma$      & 0.005 \\
$\ell_2$ weight regularization $\lambda$        & 0.42 \\
Robustness radius $\rho$                    & \{0.01, 0.1\} \\
Training epochs                             & 20 to 40 \\
Learning rate                               & $10^{-4}$ to $5\times10^{-2}$ (log-uniform) \\
Batch size                                 & 63 \\
Optimizer                                  & Adam \\
\hline
\end{tabular}
\end{table}

\begin{table}[H]
\centering
\caption{Model Configuration of SoftmaxDFL models}
\label{tab:Softmax_params}
\begin{tabular}{ll}
\hline
\textbf{Parameter} & \textbf{Values} \\
\hline
Hidden Layer                                & 32 \\
Training epochs                             & 20 to 40 \\
Learning rate                               & $10^{-4}$ to $5\times10^{-2}$ (log-uniform) \\
Batch size                                 & 63 \\
Optimizer                                  & Adam \\
\hline
\end{tabular}
\end{table}

\section{Result and Discussion}
\subsection{Overall Backtest Performance}

\begin{figure}[H] 
\centering 
\includegraphics[width=1.0\linewidth]{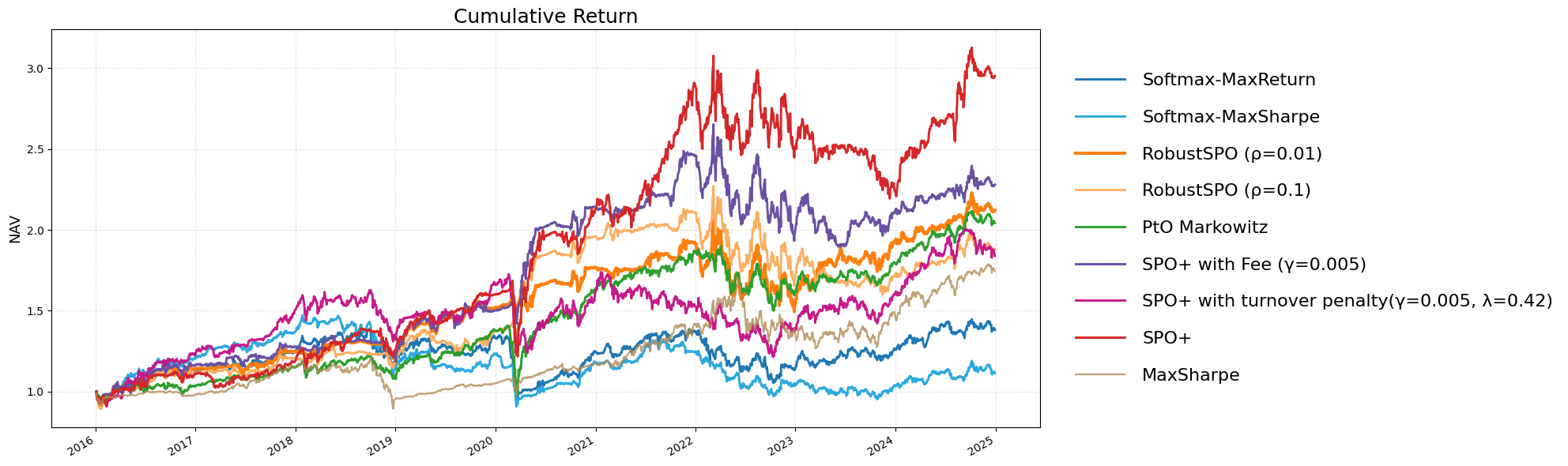} 
\caption{Cumulative net asset value (NAV) curves of all compared strategies over the full backtest period (2016--2024). } 
\label{fig:CumulativeReturn} 
\end{figure}

\begin{table}[htbp]
\centering
\caption{Overall backtest performance (2016--2024).Annualized return and volatility are reported on an annual basis.}
\label{tab:overall_backtest}
\begin{tabular}{lrrrrr}
\toprule
\textbf{Strategy} &
\textbf{Ret.} &
\textbf{Vol.} &
\textbf{Sharpe} &
\textbf{Sortino} &
\textbf{MaxDD} \\
\midrule
Softmax-MaxReturn  & 4.11  & 13.75 & 0.362 & 0.331 & -32.20 \\
Softmax-MaxSharpe  & 1.32  & 13.02 & 0.166 & 0.152 & -38.37 \\
RobustSPO ($\rho=0.01$) & \textbf{9.67}  & 15.09 & \textbf{0.688} & \textbf{0.660} & -\textbf{27.79} \\
RobustSPO ($\rho=0.1$)  & 8.09  & 14.96 & 0.595 & 0.582 & -\textbf{29.64} \\
PtO Markowitz      & 9.00  & 14.76 & 0.659 & 0.624 & -30.22 \\
SPO+ ($\gamma=0.005$) & \textbf{10.54} & 15.75 & \textbf{0.715} & \textbf{0.699} & -\textbf{28.51} \\
SPO+ ($\gamma=0.005,\ \lambda=0.42$) & 7.69 & 15.74 & 0.550 & 0.512 & -\textbf{30.18} \\
SPO+               & \textbf{14.05} & 19.09 & \textbf{0.785} & \textbf{0.728} & -\textbf{28.71} \\
MaxSharpe          & 7.13  & 13.63 & 0.574 & 0.514 & -26.33 \\
\bottomrule
\end{tabular}
\end{table}

Table~\ref{tab:overall_backtest} reports the overall backtesting performance of all compared strategies in terms of annualized return, volatility, Sharpe ratio, Sortino ratio, and maximum drawdown.

Overall, decision-focused learning approaches consistently outperform predict--then--optimize baselines and traditional portfolio optimization methods on a risk-adjusted basis.

Since the primary objective of this study is to improve \emph{decision quality} rather than predictive accuracy alone, the predict--then--optimize (PtO) framework serves as the most relevant baseline for comparison.

Among all strategies, \textbf{SPO+} achieves the highest annualized return (14.05\%) and the best risk-adjusted performance, with a Sharpe ratio of 0.785 and a Sortino ratio of 0.728.
Although SPO+ exhibits relatively higher volatility, its superior return compensates for the increased risk, resulting in the most favorable overall performance.

The \textbf{SPO+ with fee} variant also performs competitively, delivering strong risk-adjusted returns while explicitly accounting for trading costs.
In contrast, introducing the $\ell_2$ weight regularization term leads to a more conservative and diversified allocation, reducing both return and Sharpe ratio while maintaining drawdown levels comparable to other baselines.

Softmax-based strategies fail to outperform other approaches in either return or risk-related metrics.
Despite their end-to-end differentiable structure, these models do not exhibit clear advantages over predict--then--optimize or decision-focused methods in the present setting.
This performance gap may be attributed to the limited interpretability and training instability of deep neural networks when applied to noisy and non-stationary financial data, which can adversely affect the quality of the induced portfolio decisions.

\subsection{Performance during the COVID-19 Market Crisis}

\begin{figure}[H] 
\centering 
\includegraphics[width=1.0\linewidth]{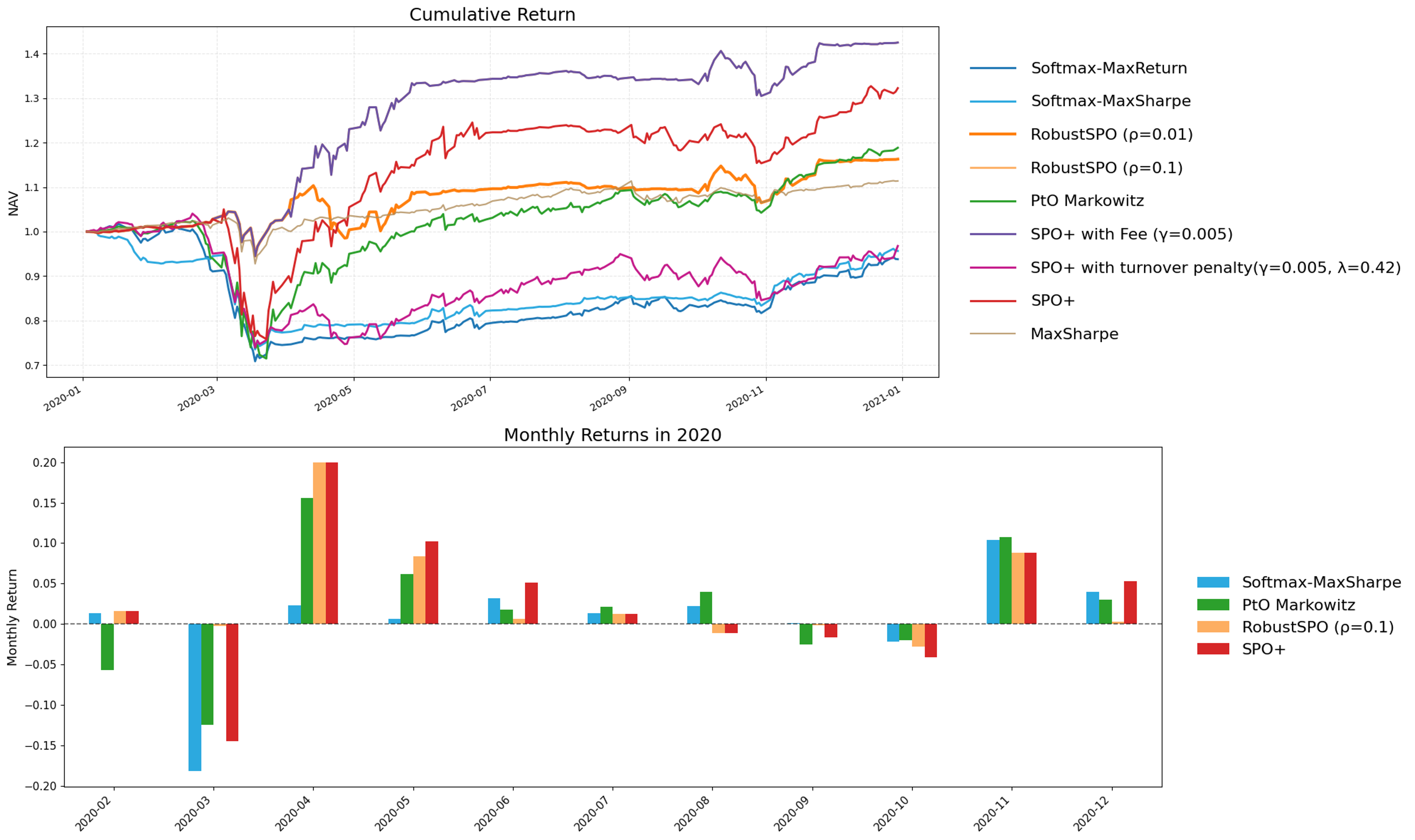} 
\caption{
Cumulative net asset value (NAV) trajectories of selected strategies
during the COVID-19 market turmoil in 2020.
The inset zooms in on the early crash period from January to April 2020.
During this interval, the \textbf{RobustSPO ($\rho=0.1$)} and
\textbf{SPO+ with Fee} strategies produce almost overlapping NAV paths,
suggesting that these models arrive at highly similar portfolio decisions
when facing severe market stress.
This behavior indicates that, under extreme market conditions, the
SPO+ with Fee solution is already sufficiently conservative due to binding constraints
and transaction cost penalties, such that additional robustness does not alter
the optimal allocation.
} 
\label{fig:2020COVID} 
\end{figure}

\begin{table}[htbp]
\centering
\caption{Backtest performance during the COVID-19 period (January 2020--December 2020).}
\label{tab:covid_backtest}
\begin{tabular}{lrrrrr}
\toprule
\textbf{Strategy} &
\textbf{Ret.} &
\textbf{Vol.} &
\textbf{Sharpe} &
\textbf{Sortino} &
\textbf{MaxDD} \\
\midrule
Softmax-MaxReturn
 & -6.73 & 18.21 & -0.292 & -0.222 & -30.37 \\
Softmax-MaxSharpe
 & -4.72 & 16.51 & -0.210 & -0.161 & -26.65 \\
RobustSPO ($\rho=0.01$)
 & \textbf{17.81} & \textbf{14.87} & \textbf{1.182} & \textbf{0.999} & \textbf{-10.71} \\
RobustSPO ($\rho=0.1$)
 & \textbf{46.92} & \textbf{18.61} & \textbf{2.170} & \textbf{2.322} & \textbf{-9.58} \\
PtO Markowitz
 & 20.63 & 28.98 & 0.796 & 0.693 & -30.22 \\
SPO+ ($\gamma=0.005$)
 & \textbf{46.92} & \textbf{18.61} & \textbf{2.170} & \textbf{2.322} & \textbf{-9.58} \\
SPO+ ($\gamma=0.005,\ \lambda=0.42$)
 & -3.51 & \textbf{22.50} & -0.046 & -0.037 & \textbf{-28.93} \\
SPO+
 & \textbf{35.48} & 31.49 & \textbf{1.128} & \textbf{0.956} & \textbf{-27.80} \\
MaxSharpe
 & 12.42 & 10.44 & 1.179 & 0.965 & -10.00 \\
\bottomrule
\end{tabular}
\end{table}

Table~\ref{tab:covid_backtest} reports the backtesting performance of all strategies during the COVID-19 period from January 2020 to December 2020, which represents an extreme market regime characterized by elevated volatility and abrupt drawdowns.
Using the predict--then--optimize (PtO) Markowitz model as the primary baseline, we highlight metrics that outperform PtO to facilitate a clear comparison in terms of decision quality.

Overall, decision-focused learning methods exhibit substantial advantages over the PtO baseline during this turbulent period.
Both \textbf{RobustSPO} variants and \textbf{SPO+ with Fee} achieve significantly higher risk-adjusted performance, as reflected by markedly improved Sharpe and Sortino ratios, while simultaneously reducing maximum drawdowns relative to PtO.
In particular, \textbf{RobustSPO ($\rho=0.1$)} and \textbf{SPO+ with Fee} demonstrate strong downside protection, with maximum drawdowns below 10\%, compared to over 30\% for the PtO baseline.

From a return perspective, several decision-focused models deliver higher annualized returns than PtO, indicating that explicitly incorporating the downstream optimization problem into the learning objective does not merely reduce risk but can also enhance profitability under stressed market conditions.
At the same time, the volatility levels of these models remain comparable to or lower than those of PtO, suggesting a more favorable risk--return trade-off.

In contrast, Softmax-based end-to-end allocation strategies perform poorly during the COVID-19 period, failing to outperform the PtO baseline across all evaluated metrics.
This observation suggests that fully differentiable allocation models without an explicit optimization structure may be more susceptible to estimation noise and training instability in highly non-stationary market environments.

Taken together, these results indicate that the benefits of decision-focused learning are particularly pronounced during adverse market regimes.
By aligning the learning objective directly with portfolio optimization outcomes, decision-focused models yield more robust and stable investment decisions than predict--then--optimize approaches when market uncertainty is elevated.

\subsection{Performance during 2024 Bull Market}

\begin{figure}[H] 
\centering 
\includegraphics[width=1.0\linewidth]{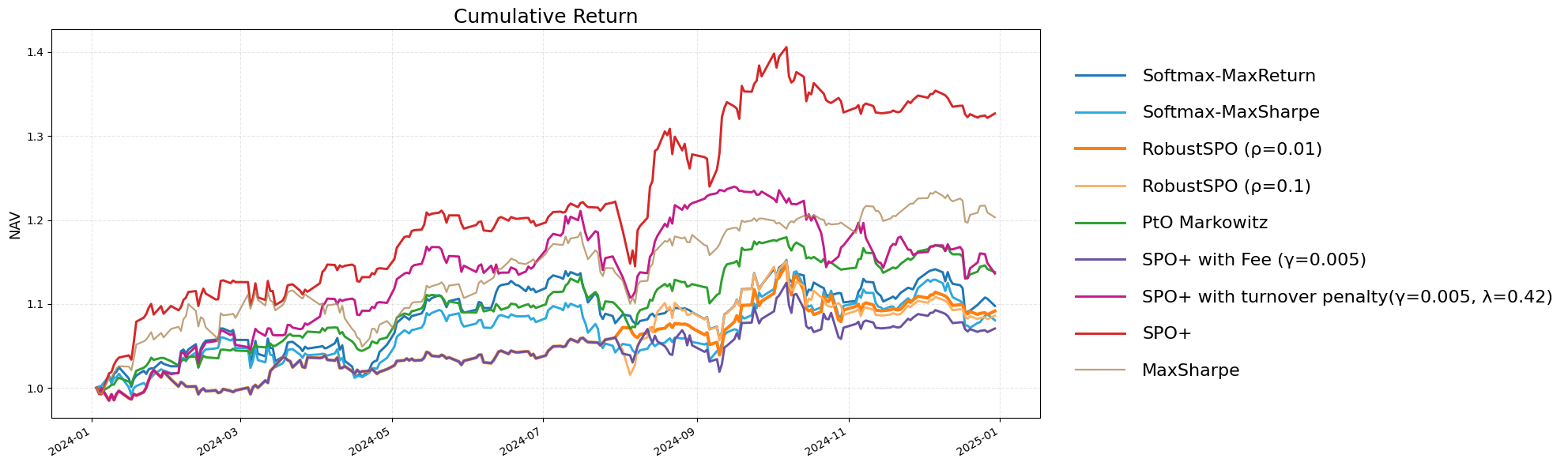} 
\caption{Cumulative net asset value (NAV) trajectories during the 2024 bull-market period.} 
\label{fig:2024Bull} 
\end{figure}

\begin{table}[htbp]
\centering
\caption{Backtest performance during the bull market period.}
\label{tab:bull_backtest}
\begin{tabular}{lrrrrr}
\toprule
\textbf{Strategy} &
\textbf{Ret.} &
\textbf{Vol.} &
\textbf{Sharpe} &
\textbf{Sortino} &
\textbf{MaxDD} \\
\midrule
Softmax-MaxReturn
 & 10.66 & 11.91 & 0.914 & 0.845 & -7.19 \\
Softmax-MaxSharpe
 & 8.79 & 12.09 & 0.760 & 0.713 & -7.02 \\
RobustSPO ($\rho=0.01$)
 & 9.99 & 9.90 & 1.015 & 0.969 & -5.94 \\
RobustSPO ($\rho=0.1$)
 & 9.38 & 10.59 & 0.903 & 0.856 & -5.99 \\
PtO Markowitz
 & 15.07 & 8.51 & 1.699 & 1.716 & -5.47 \\
SPO+ ($\gamma=0.005$)
 & 7.71 & 10.79 & 0.745 & 0.719 & -5.94 \\
SPO+ ($\gamma=0.005,\ \lambda=0.42$)
 & 14.89 & 12.72 & 1.160 & 1.092 & -8.83 \\
SPO+
 & \textbf{35.96} & 15.09 & \textbf{2.120} & \textbf{2.167} & -6.31 \\
MaxSharpe
 & 22.26 & 11.64 & 1.793 & 1.870 & -7.18 \\
\bottomrule
\end{tabular}
\end{table}

During the bull market period, models that explicitly incorporate turnover penalties or robustness constraints exhibit noticeably lower returns than the PtO Markowitz baseline.
This outcome is likely attributable to the more conservative investment behavior induced by these modeling choices, which restrains aggressive reallocation and leverage during sustained upward market trends.

This observation suggests that a substantial portion of PtO’s profitability is driven by its strong exposure to bullish market conditions.
However, it is noteworthy that the baseline \textbf{SPO+} model still outperforms the PtO baseline in several key metrics, including annualized return and risk-adjusted performance.

This result indicates that, even without additional robustness or turnover constraints, decision-focused learning can effectively capture upside opportunities while maintaining a favorable risk--return trade-off.
In this sense, \textbf{SPO+} strikes a balance between aggressiveness and stability, benefiting from bullish market dynamics without fully sacrificing robustness to market uncertainty.

\section{Conclusion}

In this study, we investigate decision-focused learning for portfolio optimization under a realistic rolling-window backtesting framework.
The empirical results demonstrate that aligning the learning objective with downstream portfolio decisions leads to consistently improved decision quality across different market regimes, compared with conventional predict--then--optimize approaches based on the Markowitz framework.

Our analysis highlights a fundamental trade-off between aggressiveness and robustness in portfolio construction.
While predict--then--optimize methods, including traditional Markowitz-based strategies, tend to benefit more from sustained bullish trends through stronger exposure to return forecasts, decision-focused models place greater emphasis on stability and downside risk control, resulting in superior performance under adverse market conditions.
Notably, the baseline SPO+ model achieves a balanced behavior across market regimes, outperforming traditional PtO baselines in several key metrics while maintaining the ability to capture upside opportunities without fully sacrificing robustness to market uncertainty.

These findings underscore the importance of incorporating decision awareness into learning-based portfolio optimization, particularly in environments characterized by noisy and non-stationary financial data, where purely prediction-driven Markowitz-style approaches may be insufficient to ensure robust decision quality.

\section{Future Work}
Several promising directions remain for future research. First, while the proposed framework accounts for decision-aware optimization under predicted returns, incorporating more advanced downside risk control mechanisms—such as distributionally robust optimization and Value-at-Risk (VaR)–type constraints may further enhance robustness against extreme market movements.  Second, although the current study adopts simple and stable predictive models, future work may explore the integration of tree-based predictors within the decision-focused learning framework~\cite{SPoTs2020}. From an empirical finance perspective, tree-based ensemble models have been shown to exhibit strong robustness in noisy return prediction tasks. For example, Krauss et al.~\cite{KRAUSS2017Review1} demonstrate that random forests and gradient-boosted trees achieve competitive and often superior out-of-sample performance compared to deep neural networks in statistical arbitrage settings, where financial signals are characterized by low signal-to-noise ratios. This robustness makes tree-based predictors a promising direction for future integration into decision-focused portfolio optimization frameworks.  Finally, despite the inclusion of diversification-promoting mechanisms, achieving an optimal balance between portfolio diversification and return maximization remains an open challenge. Developing principled approaches to jointly model this trade-off within decision-focused portfolio optimization frameworks constitutes an important direction for future research.

\newpage

\bibliography{ref}

@article{markowitz1952portfolio,
  author  = {Markowitz, Harry},
  title   = {Portfolio Selection},
  journal = {The Journal of Finance},
  volume  = {7},
  number  = {1},
  pages   = {77--91},
  year    = {1952}
}

@article{sharpe1998sharpe,
  author  = {Sharpe, William F.},
  title   = {The Sharpe Ratio},
  journal = {Journal of Portfolio Management},
  volume  = {3},
  number  = {3},
  pages   = {169--185},
  year    = {1998}
}

@article{Rockafellar2000Cvar,
  author  = {Rockafellar, R. Tyrrell and Uryasev, Stanislav},
  title   = {Optimization of Conditional Value-at-Risk},
  journal = {Journal of Risk},
  volume  = {3},
  pages   = {21--41},
  year    = {2000}
}

@article{goldfarb2003robust,
  author  = {Goldfarb, D. and Iyengar, G.},
  title   = {Robust Portfolio Selection Problems},
  journal = {Mathematics of Operations Research},
  volume  = {28},
  number  = {1},
  pages   = {1--38},
  year    = {2003},
  doi     = {10.1287/moor.28.1.1.14260}
}

@article{Jagannathan2003constraints,
  author  = {Jagannathan, Ravi and Ma, Tongshu},
  title   = {Risk Reduction in Large Portfolios: Why Imposing the Wrong Constraints Helps},
  journal = {The Journal of Finance},
  volume  = {58},
  number  = {4},
  pages   = {1651--1683},
  year    = {2003},
  doi     = {10.1111/1540-6261.00580}
}

@article{Boyd2017MultiPeriodTV,
  author  = {Boyd, Stephen P. and Busseti, Enzo and Diamond, Steven and Kahn, Ronald N. and Koh, Kwangmoo and Nystrup, Peter and Speth, Jan},
  title   = {Multi-Period Trading via Convex Optimization},
  journal = {Foundations and Trends in Optimization},
  volume  = {3},
  pages   = {1--76},
  year    = {2017}
}

@article{J.Moody2001ReccurentRL,
  author  = {Moody, J. and Saffell, M.},
  title   = {Learning to Trade via Direct Reinforcement},
  journal = {IEEE Transactions on Neural Networks},
  volume  = {12},
  number  = {4},
  pages   = {875--889},
  year    = {2001},
  doi     = {10.1109/72.935097}
}

@article{Jiang2017DRL,
  author  = {Jiang, Zhengyao and Xu, Dixing and Liang, Jinjun},
  title   = {A Deep Reinforcement Learning Framework for the Financial Portfolio Management Problem},
  journal = {arXiv preprint arXiv:1706.10059},
  year    = {2017}
}

@article{Deng2017RepresentDRL,
  author  = {Deng, Yue and Bao, Feng and Kong, Youyong and Ren, Zhiquan and Dai, Qionghai},
  title   = {Deep Direct Reinforcement Learning for Financial Signal Representation and Trading},
  journal = {IEEE Transactions on Neural Networks and Learning Systems},
  volume  = {28},
  number  = {3},
  pages   = {653--664},
  year    = {2017},
  doi     = {10.1109/TNNLS.2016.2522401}
}

@inproceedings{Pigorsch2022DQL,
  author    = {Pigorsch, Uta and Schäfer, Sebastian},
  title     = {High-Dimensional Stock Portfolio Trading with Deep Reinforcement Learning},
  booktitle = {Proceedings of the IEEE Symposium on Computational Intelligence for Financial Engineering and Economics (CIFEr)},
  pages     = {1--8},
  year      = {2022},
  doi       = {10.1109/CIFEr52523.2022.9776121}
}

@article{Liu2024PPO,
  author  = {Liu, Xiao-Yang and Xia, Ziyi and Yang, Hongyang and Gao, Jiechao and Zha, Daochen and Zhu, Ming and Wang, Christina Dan and Wang, Zhaoran and Guo, Jian},
  title   = {Dynamic Datasets and Market Environments for Financial Reinforcement Learning},
  journal = {Machine Learning},
  year    = {2024}
}

@article{Bai2025RLReview,
  author  = {Bai, Yahui and Gao, Yuhe and Wan, Runzhe and Zhang, Sheng and Song, Rui},
  title   = {A Review of Reinforcement Learning in Financial Applications},
  journal = {Annual Review of Statistics and Its Application},
  volume  = {12},
  pages   = {209--232},
  year    = {2025},
  doi     = {10.1146/annurev-statistics-112723-034423}
}

@article{Freitas2009PredictionBasedPtO,
  author  = {Freitas, Fabio D. and De Souza, Alberto F. and de Almeida, Ailson R.},
  title   = {Prediction-Based Portfolio Optimization Model Using Neural Networks},
  journal = {Neurocomputing},
  volume  = {72},
  number  = {10--12},
  pages   = {2155--2170},
  year    = {2009},
  doi     = {10.1016/j.neucom.2008.08.019}
}

@article{KRAUSS2017Review1,
  author  = {Krauss, Christopher and Do, Xuan Anh and Huck, Nicolas},
  title   = {Deep Neural Networks, Gradient-Boosted Trees, Random Forests: Statistical Arbitrage on the S\&P 500},
  journal = {European Journal of Operational Research},
  volume  = {259},
  number  = {2},
  pages   = {689--702},
  year    = {2017},
  doi     = {10.1016/j.ejor.2016.10.031}
}

@article{MA2021Review2,
  author  = {Ma, Yilin and Han, Ruizhu and Wang, Weizhong},
  title   = {Portfolio Optimization with Return Prediction Using Deep Learning and Machine Learning},
  journal = {Expert Systems with Applications},
  volume  = {165},
  pages   = {113973},
  year    = {2021},
  doi     = {10.1016/j.eswa.2020.113973}
}

@article{WANG202011LSTMforPtO,
  author  = {Wang, Wuyu and Li, Weizi and Zhang, Ning and Liu, Kecheng},
  title   = {Portfolio Formation with Preselection Using Deep Learning from Long-Term Financial Data},
  journal = {Expert Systems with Applications},
  volume  = {143},
  pages   = {113042},
  year    = {2020},
  doi     = {10.1016/j.eswa.2019.113042}
}

@article{PAIVA2019SVMforPtO,
  author  = {Paiva, Felipe Dias and Cardoso, Rodrigo Tomás Nogueira and Hanaoka, Gustavo Peixoto and Duarte, Wendel Moreira},
  title   = {Decision-Making for Financial Trading: A Fusion Approach of Machine Learning and Portfolio Selection},
  journal = {Expert Systems with Applications},
  volume  = {115},
  pages   = {635--655},
  year    = {2019},
  doi     = {10.1016/j.eswa.2018.08.003}
}

@article{DengMin2013LinearforPtO,
  author  = {Deng, Shijie and Min, Xinyu},
  title   = {Applied Optimization in Global Efficient Portfolio Construction Using Earning Forecasts},
  journal = {The Journal of Investing},
  volume  = {22},
  number  = {4},
  pages   = {104--114},
  year    = {2013},
  doi     = {10.3905/joi.2013.22.4.104}
}

@book{Deboeck1994difficulty,
  author    = {Deboeck, Guido J.},
  title     = {Trading on the Edge: Neural, Genetic, and Fuzzy Systems for Chaotic Financial Markets},
  publisher = {Wiley},
  year      = {1994}
}

@inproceedings{Zeng2021DLinear,
  author    = {Zeng, Ailing and Chen, Muxi and Zhang, Lei and Xu, Qiang},
  title     = {Are Transformers Effective for Time Series Forecasting?},
  booktitle = {Proceedings of the AAAI Conference on Artificial Intelligence},
  year      = {2023},
  doi       = {10.1609/aaai.v37i9.26317}
}

@article{elmachtoub2022SPO,
  author  = {Elmachtoub, Adam N. and Grigas, Paul},
  title   = {Smart ``Predict, then Optimize''},
  journal = {Management Science},
  volume  = {68},
  number  = {1},
  pages   = {9--26},
  year    = {2022},
  doi     = {10.1287/mnsc.2020.3922}
}

@article{mandi2024DFLReview,
  author  = {Mandi, Jayanta and Kotary, James and Berden, Senne and Mulamba, Maxime and Bucarey, Victor and Guns, Tias and Fioretto, Ferdinando},
  title   = {Decision-Focused Learning: Foundations, State of the Art, Benchmark and Future Opportunities},
  journal = {Journal of Artificial Intelligence Research},
  volume  = {80},
  pages   = {1623--1701},
  year    = {2024},
  doi     = {10.1613/jair.1.15320}
}

@misc{Wang2025EndToEndSPO,
  author = {Wang, Yi and Hasuike, Takashi},
  title  = {An End-to-End Portfolio Optimization Framework Based on the SPO Paradigm},
  year   = {2025},
  note   = {Preprint}
}

@article{tang2024pyepo,
  author  = {Tang, Bo and Khalil, Elias B.},
  title   = {PyEPO: A PyTorch-Based End-to-End Predict-Then-Optimize Library for Linear and Integer Programming},
  journal = {Mathematical Programming Computation},
  volume  = {16},
  pages   = {297--335},
  year    = {2024},
  doi     = {10.1007/s12532-024-00255-x}
}

@article{Zhang2020DeepPortfolio,
  title   = {Deep Learning for Portfolio Optimization},
  author  = {Zhang, Zihao and Zohren, Stefan and Roberts, Stephen},
  journal = {The Journal of Financial Data Science},
  year    = {2020},
  volume  = {2},
  number  = {4},
  pages   = {8--20},
  doi     = {10.3905/jfds.2020.1.042}
}

@article{Lindemann2021LSTMsurvey,
  title   = {A survey on long short-term memory networks for time series prediction},
  author  = {Lindemann, Benjamin and M{\"u}ller, Timo and Vietz, Hannes and Jazdi, Nasser and Weyrich, Michael},
  journal = {Procedia CIRP},
  volume  = {99},
  pages   = {650--655},
  year    = {2021},
  doi     = {10.1016/j.procir.2021.03.088}
}

@inproceedings{Wang2024TimeXer,
  title     = {TimeXer: Empowering Transformers for Time Series Forecasting with Exogenous Variables},
  author    = {Wang, Yuxuan and Wu, Haixu and Dong, Jiaxiang and Qin, Guo and Zhang, Haoran and Liu, Yong and Qiu, Yunzhong and Wang, Jianmin and Long, Mingsheng},
  booktitle = {Advances in Neural Information Processing Systems},
  year      = {2024},
  volume    = {37},
  pages     = {1--30}
}

@inproceedings{Nie2023PatchTST,
  title     = {A Time Series Is Worth 64 Words: Long-Term Forecasting with Transformers},
  author    = {Nie, Yuqi and Nguyen, Nam and Sinthong, Phanwong and Kalagnanam, Jayant},
  booktitle = {Advances in Neural Information Processing Systems},
  year      = {2023},
  volume    = {36},
  pages     = {1--22}
}

@inproceedings{Liu2023iTransformer,
  title     = {iTransformer: Inverted Transformers Are Effective for Time Series Forecasting},
  author    = {Liu, Yong and Qin, Guo and Wang, Jianmin and Long, Mingsheng},
  booktitle = {International Conference on Learning Representations},
  year      = {2023}
}

@inproceedings{Zhang2023TimeMixer,
  title     = {TimeMixer: Decomposable Multiscale Mixing for Time Series Forecasting},
  author    = {Zhang, Haoran and Wu, Haixu and Qin, Guo and Wang, Jianmin and Long, Mingsheng},
  booktitle = {Advances in Neural Information Processing Systems},
  year      = {2023},
  volume    = {36},
  pages     = {1--23}
}

@inproceedings{schutte2024robust,
  title     = {Robust Losses for Decision-Focused Learning},
  author    = {Schutte, Noah and Postek, Krzysztof and Yorke-Smith, Neil},
  booktitle = {Proceedings of the Thirty-Third International Joint Conference on
               Artificial Intelligence, {IJCAI-24}},
  publisher = {International Joint Conferences on Artificial Intelligence Organization},
  editor    = {Kate Larson},
  pages     = {4868--4875},
  year      = {2024},
  month     = {8},
  note      = {Main Track},
  doi       = {10.24963/ijcai.2024/538},
}

@InProceedings{SPoTs2020,
  title = 	 {Decision Trees for Decision-Making under the Predict-then-Optimize Framework},
  author =       {Elmachtoub, Adam N. and Liang, Jason Cheuk Nam and Mcnellis, Ryan},
  booktitle = 	 {Proceedings of the 37th International Conference on Machine Learning},
  pages = 	 {2858--2867},
  year = 	 {2020},
  editor = 	 {III, Hal Daumé and Singh, Aarti},
  volume = 	 {119},
  series = 	 {Proceedings of Machine Learning Research},
  month = 	 {13--18 Jul},
  publisher =    {PMLR},
}

\end{document}